# Thermal Decoherence and Population Transfer of MeV Channeling Electrons in Diamond

Tadas Paulauskas

State Research Institute Center for Physical Sciences and Technology, Saulėtekio al. 3, Vilnius LT-10257, Lithuania

Published in Nuclear Instruments and Methods in Physics Research B 581 (2026) 166335. DOI: 10.1016/j.nimb.2026.166335

## Abstract

Channeling radiation from MeV-regime electrons is governed by transitions between quantized transverse bound states, but experimental spectra are strongly modified by thermal diffuse scattering. To capture these open-system dynamics, a frozen-phonon multislice framework is combined with bound-state projection analysis to construct depth-dependent reduced density matrices in selected transverse manifolds. Beyond reproducing experimental channeling-radiation transition energies, this approach separates thermal population transfer, intra-manifold decoherence, and cross-manifold coherence loss. Applied to 16.9 MeV axial electron channeling in $\langle 100 \rangle$ diamond, the results show approximately exponential population decay from the initially occupied states, accompanied by strongly channel-dependent feeding among low-lying manifolds. Starting from a coherent superposition within the degenerate $2p$ manifold, stochastic symmetry breaking by thermal displacements drives the intra-manifold purity toward the maximally mixed limit, indicating rapid phase scrambling. Under $1s$ initialization, population transferred into the $2p$ and $3d$ manifolds remains internally close to maximally mixed, while the finite $2p - 3d$ cross-manifold coherence is reported here as an upper bound. This framework goes beyond static mean-field thermal broadening and provides a microscopic basis for evaluating population dynamics and coherence lifetimes in strongly quantized channeling-radiation systems.

## 1. Introduction

In the few-MeV regime, electron channeling in oriented crystals is strongly quantized, and channeling-radiation spectra are governed by transitions between discrete transverse bound states [1-3]. Experimental spectra in this regime are not determined by the static bound-state spectrum alone: coupling to lattice vibrations produces thermal diffuse scattering (TDS), line broadening, dechanneling, and state feeding during propagation [4-7]. At hundreds-MeV to GeV energies, classical trajectory descriptions are often adequate because the transverse spectrum becomes quasi-continuous [8,9]. In contrast, the lower-MeV regime requires a quantum description of the transverse motion.

Quantum band-structure approaches, including Bloch-wave and tight-binding treatments with Debye–Waller broadened mean-field potentials, accurately yield thermally shifted transverse eigenenergies and dispersive wavefunctions [6,8-13]. However, a static thermal potential cannot describe the open-system dynamics induced by thermal diffuse scattering. In particular, it does not explicitly track non-radiative population transfer between bound manifolds or the decay of off-diagonal coherence terms in the propagating transverse wavefunction. Phenomenological attenuation factors, optical potentials, and rate-equation descriptions can account for dechanneling and bulk population flow, but they generally do not provide direct access to the reduced density matrix of selected bound-state manifolds. This distinction is important because polarization, angular distributions, and coherence-sensitive channeling-radiation observables depend on the phase structure of the transverse state, not only on its population. It is also central to assessing externally driven or stimulated channeling-emission schemes, where the survival of phase correlations determines the achievable control contrast [14-18].

To go beyond static thermal broadening, this work employs a paraxial multislice propagator combined with the frozen-phonon (FP) method [19-21]. The frozen-phonon approach, widely validated in atomic-resolution transmission electron microscopy, represents thermal diffuse scattering through an ensemble of static, thermally displaced crystal configurations. Applied to channeling radiation, it retains full wave propagation while capturing configuration-dependent transverse momentum transfer, state-selective feeding, and phase scrambling. The same framework is naturally compatible with arbitrary crystal structures, including defective or bent lattices [19].

Here, the frozen-phonon multislice approach is adapted to track the quantum evolution of MeV channeling electrons in a reduced transverse Hilbert space. Configuration-dependent wavefunctions are projected onto selected bound-state manifolds during propagation, allowing extraction of depth-dependent manifold populations, reduced density matrices within degenerate subspaces, and basis-invariant cross-manifold coherence measures. The method is applied to axial electron channeling in ⟨100⟩ diamond in the 16.9 MeV case study. The analysis focuses on thermal population transfer, rapid intra-manifold decoherence in degenerate manifolds, and finite-sample cross-manifold coherence estimates for the dominant $2p \to 1s$ and $3d \to 2p$ radiative channels**.**

## 2. Theoretical formulation

The standard paraxial approximation for high-energy, spinless channeling electrons is used. The fast longitudinal motion is separated from the slowly varying transverse envelope $\psi(\boldsymbol{r}, z)$, with z acting as the propagation coordinate. The transverse envelope evolves according to the unitary propagator

(1) $$\psi(\boldsymbol{r}, z) = U(z, z_0)\, \psi(\boldsymbol{r}, z_0).$$

The crystal is discretized into $N$ atomic slices of thickness $\Delta z$ and the slice potential $V_j$ is defined as the $z$-average over the slice. The exact evolution is then written in terms of the z-ordered transverse Hamiltonians $H_\perp(z')$:

(2) $$U(z, z_0) = \mathcal{T} \exp\left[-i\, \sigma \int_{z_0}^{z} H_\perp(z')\, dz'\right].$$

Here $\sigma$ is the relativistic interaction constant. For numerical propagation, the non-commuting kinetic and potential terms of $H_\perp$ are factorized using the standard split-operator approximation [22]. The per-slice propagator $U_j$ and the wavefunction after thickness $N\Delta z$ are therefore:

(3) $$U_j = \exp\left[i\sigma(K + V_j)\Delta z\right] \cong \exp[i\sigma K\Delta z]\exp\left[i\sigma V_j\Delta z\right],$$

(4) $$\psi(\boldsymbol{r}, N\Delta z) \cong \left(\prod_{j=1}^{N} U_j\right)\psi(\boldsymbol{r}, 0).$$

Above, $K$ denotes the negative of the transverse kinetic-energy operator and $V_j$ the projected potential of slice $j$. The local error of the first-order Suzuki–Trotter factorization in eq. (3) is $\mathcal{O}(\Delta z^2)$, and convergence is controlled by reducing the slice thickness $\Delta z$ (Supplemental Material, Secs. S1 and S3, eqs. S2, S3). The crystal is constructed from the Doyle–Turner parametrization of atomic potentials [23]. In the corresponding static thermal treatment, the isotropic Debye–Waller factor with the two-dimensional mean square amplitude of thermal vibrations, $u_{2d}^2$, broadens the atomic potentials and produces a mean-field thermally smeared lattice.

Thermal diffuse scattering is treated here using the frozen-phonon approximation. In this approach, the thermal phonon statistical mixture is represented, under the harmonic approximation, by an ensemble of static lattice configurations indexed by $\beta = 1, \ldots, N\beta$. For each configuration $\beta$, atomic positions are obtained by applying random thermal displacements $\boldsymbol{u}_{\boldsymbol{l}}^{(\beta)}$ to each atom $\boldsymbol{l}$, $\boldsymbol{R}_{\boldsymbol{l}}^{(\beta)} = \boldsymbol{R}_{\boldsymbol{l}}^{(\boldsymbol{0})} + \boldsymbol{u}_{\boldsymbol{l}}^{(\beta)}$. The displacements $\boldsymbol{u}_{\boldsymbol{l}}^{(\beta)}$ are sampled from a zero-mean Gaussian distribution with variance set by $u_{2d}^2$. For a given configuration number $\beta$, the electron evolves coherently through the corresponding static displaced lattice,

(5) $$\psi^{(\beta)}(\boldsymbol{r}, z) = U^{(\beta)}(z, 0)\,\psi(\boldsymbol{r}, 0).$$

Experimentally relevant intensity-like observables and spectral densities are obtained by incoherent averaging over $\beta$, equivalent to tracing out the unobserved phonon subsystem. The channeling electron transverse wavefunction is analyzed using an eigenmode expansion,

(6) $$\begin{gathered}\psi(\boldsymbol{r}, z) \approx \sum_n c_n(z)\,\phi_n(\boldsymbol{r})\exp[-i\sigma\varepsilon_n z], \\ c_n(0) = \int d^2r\,\phi_n^*(\boldsymbol{r})\psi(\boldsymbol{r}, 0)\end{gathered}.$$

The localized, non-dispersive bound states $\phi_n$ and their transverse energies $\varepsilon_n$ are extracted using the spectral propagation method [24] (Supplemental Material, Secs. S1 and S6, eq. S7). The resulting bound-state basis is then used to compute projection amplitudes, cross-correlation functions, manifold populations, reduced density matrices, and coherence metrics as functions of propagation depth in the thermally displaced lattice ensemble.

## 3. Results

### 3.1 Axial channeling radiation transition energies

To begin, axial channeling radiation transition energies for diamond along ⟨100⟩ and ⟨110⟩ axes at several beam energies are compared against reported experimental values to validate the numerical approach [25,26]. The root mean square (RMS) two-dimensional thermal displacement amplitude, $u_{2d}$, for carbon is set to $0.055$ Å in the frozen-phonon ensemble approach.

To generate the autocorrelation function, the initial electron wavefunction was chosen as a superposition of several transverse profiles with different symmetries and overlap onto the relevant bound modes. Figure 1 shows the ensemble-averaged imaginary part of the FP autocorrelation function, $Im\langle A^{(\beta)}(z)\rangle_{\beta}$, with $\beta = 10$ phonon configurations for $\langle 100\rangle$ diamond at 16.9 MeV and 30.5 MeV. The corresponding DW atomic smearing result is also shown for comparison using the same $u_{2d}$ to parametrize it.

In the FP case, the autocorrelation exhibits a clear decay with thickness, and no comparable decay is observed over the same thickness using DW-smeared potentials. This can be explained by stochastic transverse momentum transfer under thermal atomic displacements, which are absent in a static mean-field description.

The transverse energy spectra are shown alongside the autocorrelations in Fig. 1. For the spectrum, the absolute value of an incoherent ensemble average is used, $Abs\langle|S^{(\beta)}(k_z)|\rangle_{\beta}$. An important observation is that the extracted bound-state energies (peak positions) agree, up to the numerical energy-bin size ($E_{bin}$~0.69 eV), between the FP and DW approaches. This agreement was observed across the tested beam energies for $\langle 100\rangle$ and $\langle 110\rangle$ orientations, indicating that the FP sampling reproduces the thermal diffuse effects at the level of eigen-energies. Peak widths are similar for both FP and DW cases and are set by decay rate of the bound-state as well as finite propagation depth, which imposes Fourier broadening and can dominate for weakly decaying states.

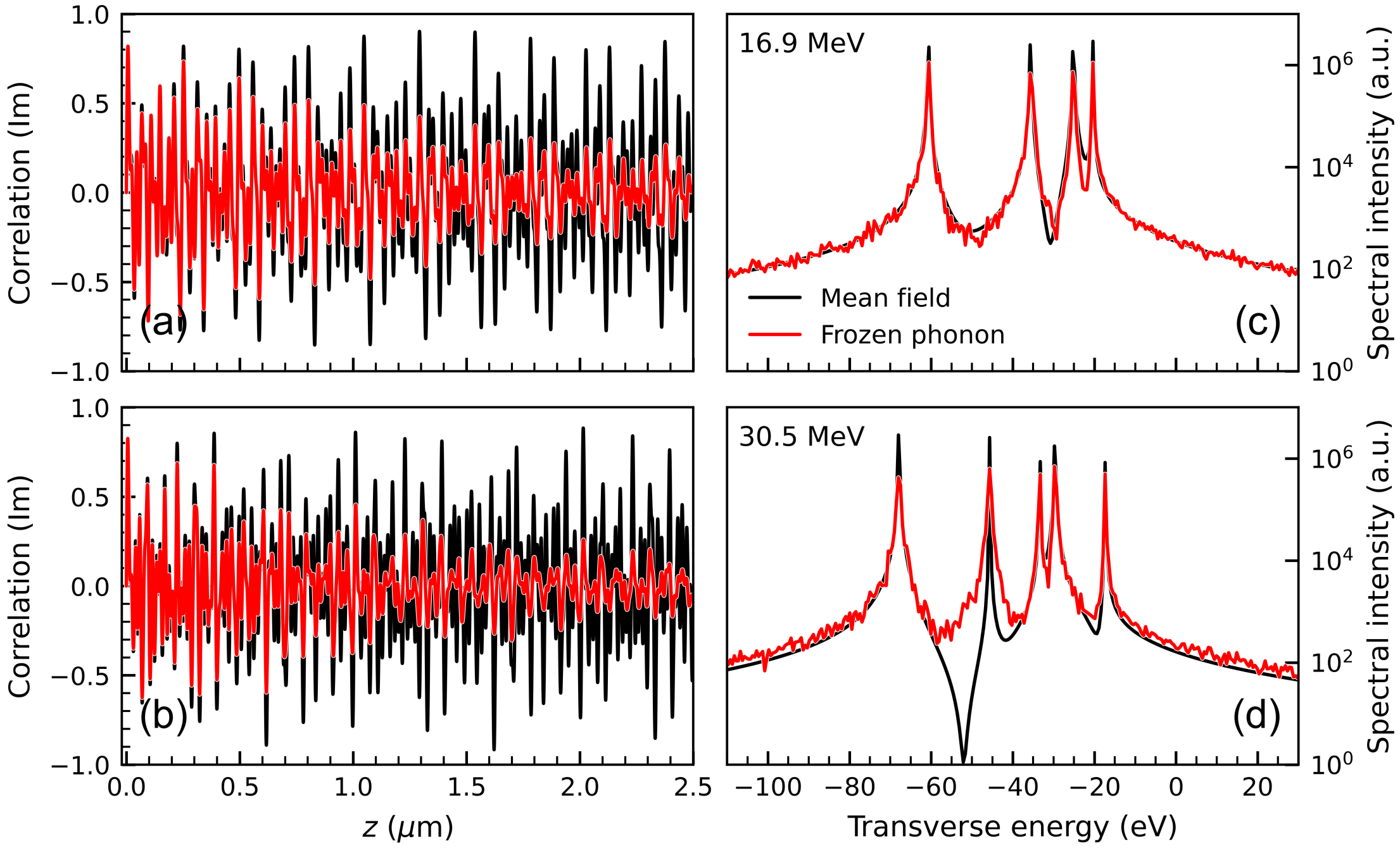

Fig. 1. (a, b) Imaginary parts of the ensemble-averaged autocorrelation functions at 16.9 MeV and 30.5 MeV beam energies. Mean-field (black) and frozen-phonon (red) calculations are overlaid. (c, d) Transverse energy spectrum plotted in its absolute value in log-scale.

The peaks in the transverse-energy spectrum in hydrogenic notation correspond to $1s$, $2p$, $2s$, $3d$, and $3p$ ($3p$ is excited here only in the 30.5 MeV case), from the deepest to shallower bound states, respectively. To relate the energy spacings $\Delta E$ of transitions to laboratory photon energies, the standard small-angle forward-emission approximation is used [1-3]:

$$E_{\gamma}(\theta) = \frac{2\gamma^2 \Delta E}{1+\gamma^2\theta^2} \quad (7)$$

Channeling radiation energies reported in Table 1 were evaluated at $\theta = 0$ and compared against experimentally quoted transition values for several beam energies and crystal orientations [25,26]. It shows high agreement across most of the axial transitions, supporting this approach. The largest discrepancy is observed for the $3d \rightarrow 2p$ transition at 30.5 MeV, where the difference exceeds that of other transitions and may require closer scrutiny. Because the $3d$ state is weakly bound and spatially extended, it is more sensitive to the continuum-potential parametrization. Higher-order thermal processes or non-ideal crystal effects not included in the present ideal-crystal model may also contribute.

The transition energies are sensitive to the thermal displacement amplitude, particularly for the more deeply bound states. Values for $u_{2d}$ in the range $0.050 - 0.070$ Å are commonly found in the literature when modelling thermal linewidth broadening, and the adopted value here ($0.055$ Å) provides a representative match to the experimental transition energies [6,25]. Additional dependence of transverse eigenvalues at 16.9 MeV on the $u_{2d}$ is provided in Supplemental Material, Sec. S4.

Table 1. A comparison of experimentally reported and in this work calculated axial channeling radiation transition energies in diamond [25,26]. Uncertainty values provided in the observed literature values in parentheses, where available.

| **Beam energy** | **Axial transition** | **Observed energy (keV)** | **Calculated (keV)** |
|---|---|---|---|
| 4 MeV [25] | ⟨100⟩, 2p–1s | 3.80 | 3.767 |
| 16.9 MeV [25] | ⟨100⟩, 2p–1s | 58.3(0.5) | 58.48 |
| | ⟨100⟩, 3d–2p | 35.0(0.5) | 35.288 |
| 30.5 MeV [25] | ⟨100⟩, 2p–1s | 161.6(2) | 163.17 |
| | ⟨100⟩, 3d–2p | 110.4(1) | 118.381 |
| 5.2 MeV [26] | ⟨110⟩, 2p–1s | 9.38(0.09) | 8.804 |
| 9.0 MeV [26] | ⟨110⟩, 2p–1s | 24.34(0.23) | 24.481 |

### 3.2 Population dynamics under frozen-phonon scattering

To investigate population dynamics and coherence properties, a case study is done first for axial channeling in diamond along ⟨100⟩ at 16.9 MeV electron beam energy (and the same 0.055 Å $u_{2d}$ value). Using the observation that eigenenergies between the FP and DW approaches agree for the same $u_{2d}$ values, eigenmodes are extracted from the DW-broadened crystal, and then employed as initial (eigen)states for propagation within a crystal with frozen-phonon atomic displacements and for computation of relevant correlation functions. Numerical propagation is carried out up to 2.8 $\mu m$ crystal thickness with $N_\beta = 30$ phonon configurations. Relevant notation and the extracted ensemble quantities are described next. Convergence with respect to the number of configurations is provided in Supplemental Material, Sec. S2.

The transverse dynamics are analyzed in a tracked Hilbert space decomposed into bound-state manifolds. In an ideal cylindrically symmetric column potential, these manifolds can be indexed by radial and azimuthal quantum numbers, $(n, m)$ [13]. Here, the corresponding hydrogenic labels are used for brevity. Each manifold $M \in \{1s, 2s, 2p, 3d\}$ defines a subspace $\mathcal{H}_M$ spanned by eigenmodes $\{|M, a\rangle\}$ of the transverse Hamiltonian. The $1s$ and $2s$ manifolds are non-degenerate, while $2p$ and $3d$ are twofold degenerate ($\dim \mathcal{H}_M = 2$). The letter $a$ in $|M, a\rangle$ labels a particular basis state of a chosen basis in the degenerate manifold $M$.

The evolution of the electron under frozen-phonon scattering is captured by projecting the phonon configuration-dependent wavefunction $\psi^{(\beta)}(\boldsymbol{r}, z)$ onto the basis states:

$$(8) \quad c_{Ma}^{(\beta)}(z) = \langle M, a | \psi^{(\beta)}(z) \rangle.$$

The total population in manifold $M$ for configuration $\beta$ is basis-invariant Born probability:

(9) $p_M^{(\beta)}(z) = \sum_a \left| c_{Ma}^{(\beta)}(z) \right|^2,$

To quantify population transfer, the incoherent conditional population is defined as:

(10) $P_M(z|M_0) = \left\langle p_M^{(\beta)}(z) \right\rangle_\beta,$

which represents the ensemble-averaged population in manifold $M$ at depth $z$ given an initial excitation in manifold $M_0$.

To obtain Fig. 2, the propagation is initialized in a selected normalized transverse eigenmode $\psi(0) = |M_0, a_0\rangle$ of one manifold $M_0 \in \{1s, 2s, 2p, 3d\}$, and projections are tracked into the set of manifolds $M = \{1s, 2s, 2p, 3d\}$ as a function of depth $z$. Shaded bands indicate one standard deviation variations across frozen-phonon configurations $\beta$ at each $z$. Figure 2 shows that, for all initializations, the population in the originating manifold $M_0$ decays approximately exponentially under thermal diffuse scattering. More tightly bound states depopulate more rapidly, which is expected for their increased sensitivity to atomic displacements. Population transfer among the tracked manifolds is distinctly channel dependent. The strongest bound-to-bound coupling occurs between the $1s$ and $2p$, which, when starting from zero-population, display a clear rise due to feeding followed by decay. Coupling between $2p$ and $3d$ is also substantial (panels c–d), whereas $2s$ couples comparatively weakly to the other tracked manifolds.

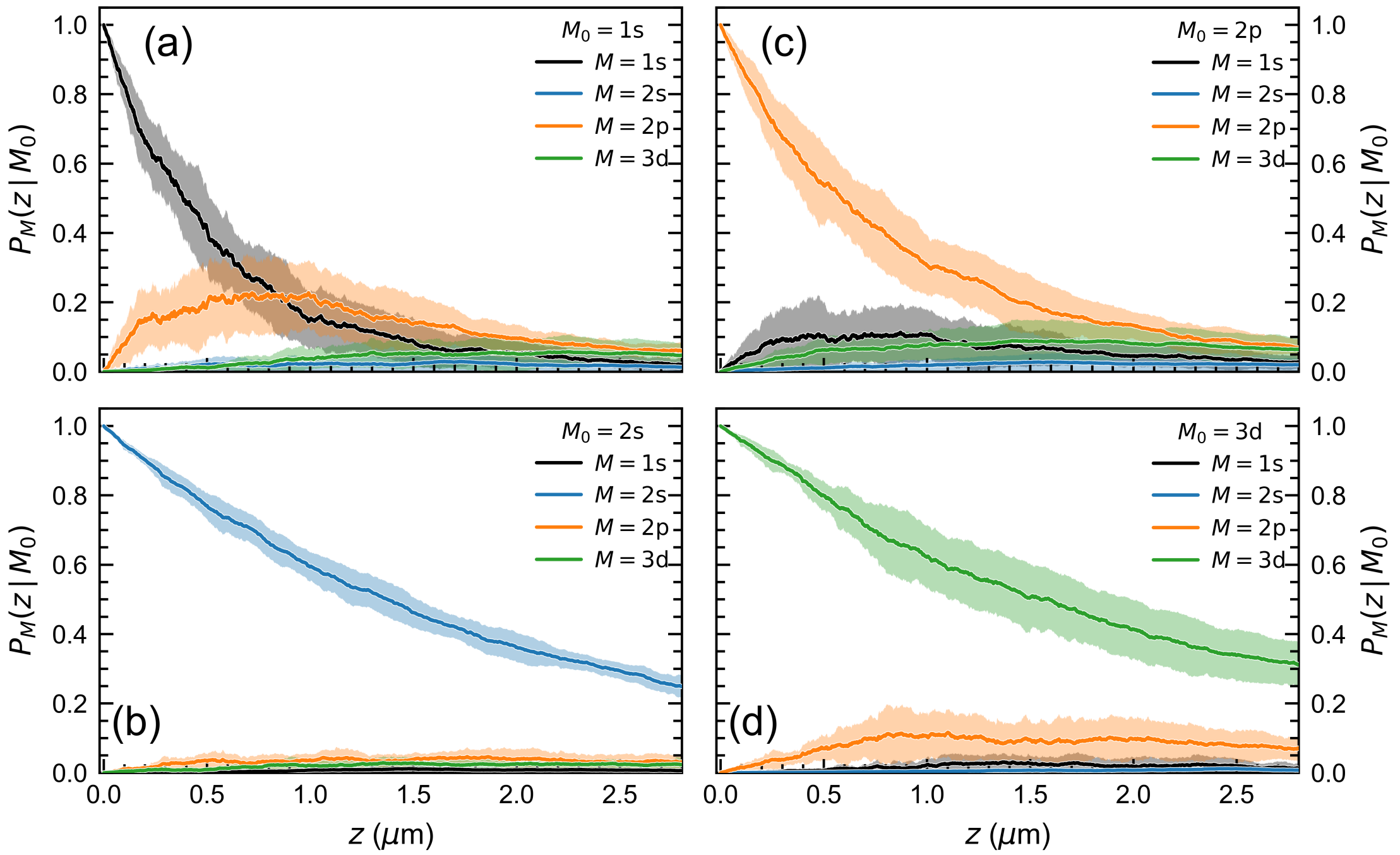

Fig. 2. (a-d) Ensemble-averaged manifold populations $P_M(z|M_0)$ in the tracked bound-state manifolds $\mathcal{M} = \{1s, 2s, 2p, 3d\}$ as a function of depth $z$, for 16.9 MeV axial channeling in $\langle 100 \rangle$ diamond. Each panel corresponds to propagation initialized in an eigenmode of manifold $M_0$, and the curves show the population transferred into each manifold $M$. Shaded regions indicate $\pm 1\sigma$ variations across frozen-phonon configurations.

### 3.3 Coherence within a degenerate manifold

Two coherence questions are considered in this work: (i) intra-manifold coherence, describing phase relations within a degenerate bound-state manifold, and (ii) cross-manifold coherence, describing phase relations between distinct manifolds. Both analyses below use 16.9 MeV channeling in $\langle 100 \rangle$ diamond. The corresponding manifold populations $P_M(z|M_0)$ are shown in Fig. 2.

This section focuses on coherence in the $2p$ manifold. For each frozen-phonon configuration $\beta$, the propagated wavefunction is projected onto a chosen orthonormal basis spanning the $2p$ manifold. An orbital-angular-momentum (OAM) basis $\{|+\rangle, |-\rangle\} \equiv \{|m = +1\rangle, |m = -1\rangle\}$ is chosen and the initial state is prepared as their equal superposition:

$$(11)\quad \left|\psi^{(\beta)}(0)\right\rangle = |2p, +x\rangle \equiv \frac{|+\rangle + |-\rangle}{\sqrt{2}},$$

i.e. the $+1$ eigenstate of $\sigma_1$ in the $\{|+\rangle, |-\rangle\}$ basis. For each frozen-phonon configuration $\beta$, the $2p$ projection amplitudes are collected into the two-vector:

$$(12)\quad \boldsymbol{c}^{(\beta)}(z) = \left(c_+^{(\beta)}(z), c_-^{(\beta)}(z)\right)^T,$$

$$(13)\quad c_\pm^{(\beta)}(z) = \langle \pm | \psi^{(\beta)}(z) \rangle.$$

The corresponding $2p$ manifold population is thus:

$$(14)\quad p_{2p}^{(\beta)}(z) = \left|c_{+}^{(\beta)}(z)\right|^2 + \left|c_{-}^{(\beta)}(z)\right|^2.$$

To separate internal coherence from overall population loss, the projected state is normalized within that manifold:

$$(15)\quad \boldsymbol{u}^{(\beta)}(z) = \frac{\boldsymbol{c}^{(\beta)}(z)}{\sqrt{p_{2p}^{(\beta)}(z)}},$$

which is well-defined when $p_{2p}^{(\beta)}(z) > \varepsilon$. Configurations below a chosen threshold value, $p_{2p}^{(\beta)}(z) \le \varepsilon$, are omitted from the ensemble average calculations at that depth. The ensemble-averaged reduced density matrix (trace-1) within the $2p$ manifold, conditioned on occupying $2p$ at depth $z$, is then:

$$(16)\quad \bar{\rho}_{2p}(z) = \left\langle \boldsymbol{u}^{(\beta)}(z)\boldsymbol{u}^{(\beta)}(z)^{\dagger}\right\rangle_{\beta}.$$

Here and below, the overbar denotes averaging over frozen-phonon configurations after normalization to unit trace within the selected manifold. For each configuration $\beta$, the Bloch components are computed from the normalized two-vector as:

$$(17)\quad S_i^{(\beta)}(z) = \boldsymbol{u}^{(\beta)}(z)^{\dagger}\,\sigma_i\,\boldsymbol{u}^{(\beta)}(z),$$

where $\sigma_i$ are the Pauli matrices. In the $\{|+\rangle, |-\rangle\}$ basis this gives:

$$(18)\quad S_1^{(\beta)}(z) = \frac{2\mathrm{Re}\left(c_{+}^{(\beta)}(z)c_{-}^{(\beta)}(z)^{*}\right)}{p_{2p}^{(\beta)}(z)},$$

$$(19)\quad S_2^{(\beta)}(z) = \frac{-2\mathrm{Im}\left(c_{+}^{(\beta)}(z)c_{-}^{(\beta)}(z)^{*}\right)}{p_{2p}^{(\beta)}(z)},$$

$$(20)\quad S_3^{(\beta)}(z) = \frac{|c_{+}^{(\beta)}(z)|^2 - |c_{-}^{(\beta)}(z)|^2}{p_{2p}^{(\beta)}(z)}.$$

The $2p$ manifold purity is obtained from:

$$(21)\quad \gamma(z) = \mathrm{Tr}\left[\bar{\rho}_{2p}(z)^2\right],$$

which is basis invariant and $1/2 \le \gamma(z) \le 1$. As a basis-dependent metric, the coherence of the ensemble-averaged state is also reported, $0 \le \kappa(z) \le 1$, obtained from the normalized $\bar{\rho}_{2p}(z)$ off-diagonal

$$(22)\quad \kappa(z) = 2\left|\bar{\rho}_{2p,+-}(z)\right|.$$

Figure 3(a) shows $\gamma(z)$ together with $\kappa(z)$ for the initially pure superposition ($\gamma(0) = \kappa(0) = 1$). The purity decreases with depth toward the maximally mixed limit ($\gamma \to 1/2$), while $\kappa(z)$ decays to small values. Figure 3(b) shows the ensemble-averaged Bloch components $\langle S_i(z)\rangle_{\beta}$. The component aligned with the prepared superposition, $\langle S_1(z)\rangle_\beta$, starts near unity and decays with depth, whereas $\langle S_2(z)\rangle_\beta$ and the population-imbalance component $\langle S_3(z)\rangle_\beta$ remain near zero in the ensemble mean. In addition, the $\pm1\sigma$ bands of $\langle S_i(z)\rangle_\beta$ show significant configuration-to-configuration dispersion of Bloch trajectories.

These results, taken together, are indicative of stochastic and $\beta$ -dependent internal-basis rotations within the degenerate manifold. In a fixed basis employed for the analysis, this appears as the relative-phase randomization and rapid statistical mixing within the manifold. The same behavior is expected in any internal basis of a degenerate manifold, since the ensemble of mean-zero, approximately isotropic thermal displacements does not select a preferred axis. The analysis was also performed in a rotated $2p$ basis ($|\pm x\rangle$), preparing an initial state with $\langle S_3(0)\rangle_\beta = 1$. Again, the initially nonzero Bloch component decays toward zero and $\gamma(z) \to 1/2$. Applying the same diagnostics to the degenerate $3d$ manifold (not shown) reveals substantially slower mixing, consistent with its more spatially extended transverse modes and reduced sensitivity to thermal atomic displacements.

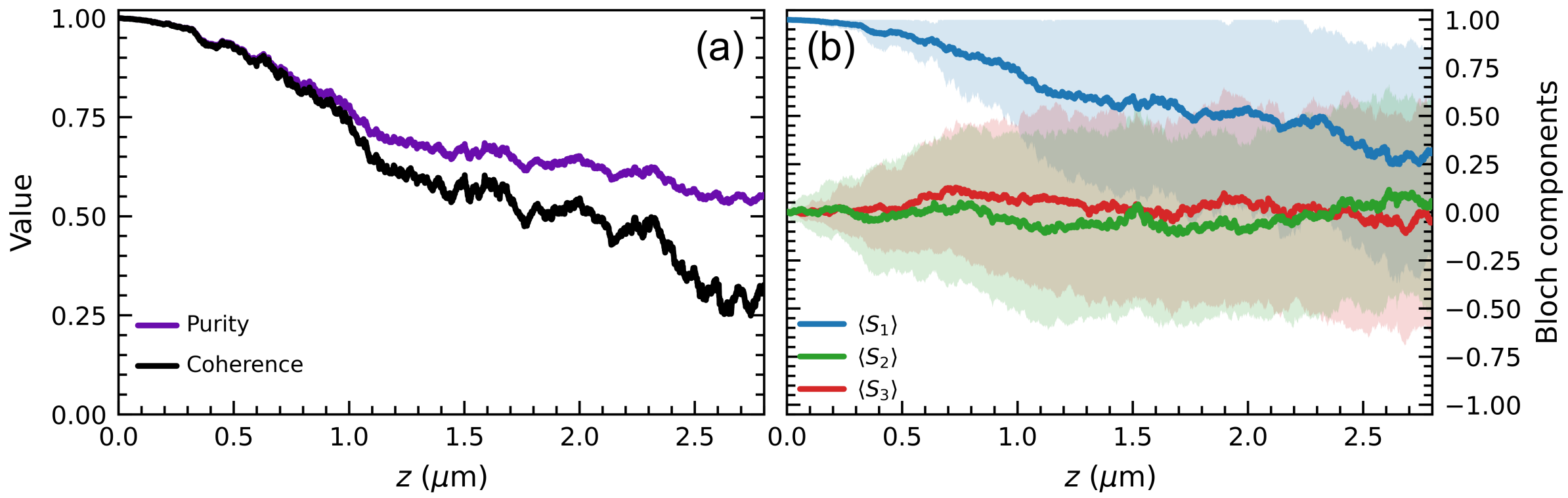

Fig. 3. (a) Ensemble-averaged purity $\gamma(z)$ and ensemble-averaged normalized coherence $\kappa(z)$, in the $M = 2p$ manifold for initial $M_0 = 2p$ excitation starting from an equal superposition of the two OAM basis states $|+\rangle$ and $|-\rangle$. (b) Ensemble-averaged Bloch components $\langle S_i(z)\rangle_\beta$ with shaded $\pm1\sigma$ bands across configurations $\beta$.

### 3.4 Cross-manifold coherences

In on-axis channeling radiation experiments the incident beam is typically collimated or weakly convergent, illuminating many atomic columns simultaneously. Among bound states, the initial overlap is therefore dominated by the $m = 0$ manifolds. Population in $|m| > 0$ manifolds, and therefore the radiative channels that involve them, then arises primarily through non-radiative, quasi-elastic thermal diffuse scattering. This section quantifies the depth-resolved purity of the initially unoccupied $2p$ and $3d$ manifolds and the surviving cross-manifold coherence in the radiative channels $2p \to 1s$ and $3d \to 2p$. For this, the same 16.9 MeV channeling in $\langle 100\rangle$ diamond setting but with an initially prepared $1s$ eigenmode is employed.

To quantify coherence between two manifolds $A$ and $B$ with dimensions $d_A, d_B \in \{1,2\}$, we first collect the per-configuration projection amplitudes as two-vectors, $\boldsymbol{a}^{(\beta)}(z) \in \mathbb{C}^{d_A}$ and $\boldsymbol{b}^{(\beta)}(z) \in \mathbb{C}^{d_B}$, in chosen orthonormal bases within each manifold. The associated cross-block matrix is:

$$(23) \quad \boldsymbol{X}_{AB}^{(\beta)}(z) = \boldsymbol{a}^{(\beta)}(z)\boldsymbol{b}^{(\beta)}(z)^{\dagger} \in \mathbb{C}^{d_A \times d_B}.$$

Using ensemble-averaged cross block, $\langle \boldsymbol{X}_{AB}(z)\rangle_\beta$, a basis-invariant Frobenius-normalized cross-coherence, $0 \leq C_F^{A,B}(z) \leq 1$, is defined as:

$$(24) \quad C_F^{A,B}(z) = \frac{\left\| \langle X_{AB}(z)\rangle_\beta \right\|_F}{\sqrt{\langle p_A(z)\rangle_\beta \langle p_B(z)\rangle_\beta}},$$

$$(25) \quad p_A^{(\beta)}(z) = \left\|\boldsymbol{a}^{(\beta)}(z)\right\|^2, p_B^{(\beta)}(z) = \left\|\boldsymbol{b}^{(\beta)}(z)\right\|^2.$$

The Frobenius norm, $\|\cdot\|_F$, ensures invariance of $C_F^{A,B}(z)$ under independent unitary basis rotations within manifolds $A$ and $B$ [27]. Numerically, $C_F^{A,B}(z)$ is evaluated only where the denominator exceeds a threshold to avoid instabilities at vanishing populations. In practice, the number of contributing FP configurations remains near maximal ($\sim N_\beta$) across the depth range, indicating that the values are not biased toward a small subset.

Purities in the $2p$ and $3d$ manifolds are evaluated as in the previous section, but here for initialized pure $1s$ eigenmode. Figure 4(a) shows that the conditioned intra-manifold purities of $2p$ and $3d$ remain close to the maximally mixed limit, $\gamma(z) \approx 1/2$, over the full depth range. This indicates that population transferred from $1s$ into these degenerate manifolds does not retain coherence in any fixed intra-manifold basis. The inset in Fig. 4(a) confirms this showing the ensemble-averaged Bloch components in $2p$ remain near zero with large configuration-to-configuration spread. The same behavior is also observed for $3d$ (not shown).

Figure 4(a) shows the Frobenius-normalized cross-coherences for the manifold pairs $(1s, 2p)$ and $(2p, 3d)$ under $1s$ excitation. As can be seen, the cross-coherence takes a small finite value with $C_F^{A,B}(z) \sim 0.2$ over the depth range. However, its bootstrap interval overlaps a shuffled-configuration null and hence is not statistically distinguishable from finite-ensemble bias (see Supplemental Material, Sec. S5). Therefore, $C_F^{A,B}(z) \sim 0.2$ is presently interpreted as the upper bound. Fig. 4(b) shows the (population-unnormalized) real parts of the individual ensemble-averaged cross-block elements $\mathrm{Re}\,\langle X_{ij}(z)\rangle_\beta$ for the $2p-3d$ block, with $i$ and $j$ indexing the chosen bases in $2p$ and $3d$. Each element exhibits a small oscillatory component after ensemble averaging, with frequency set by the transverse eigenenergy differences. The oscillation amplitudes and relative phases drift with depth, reflecting stochastic thermal perturbations.

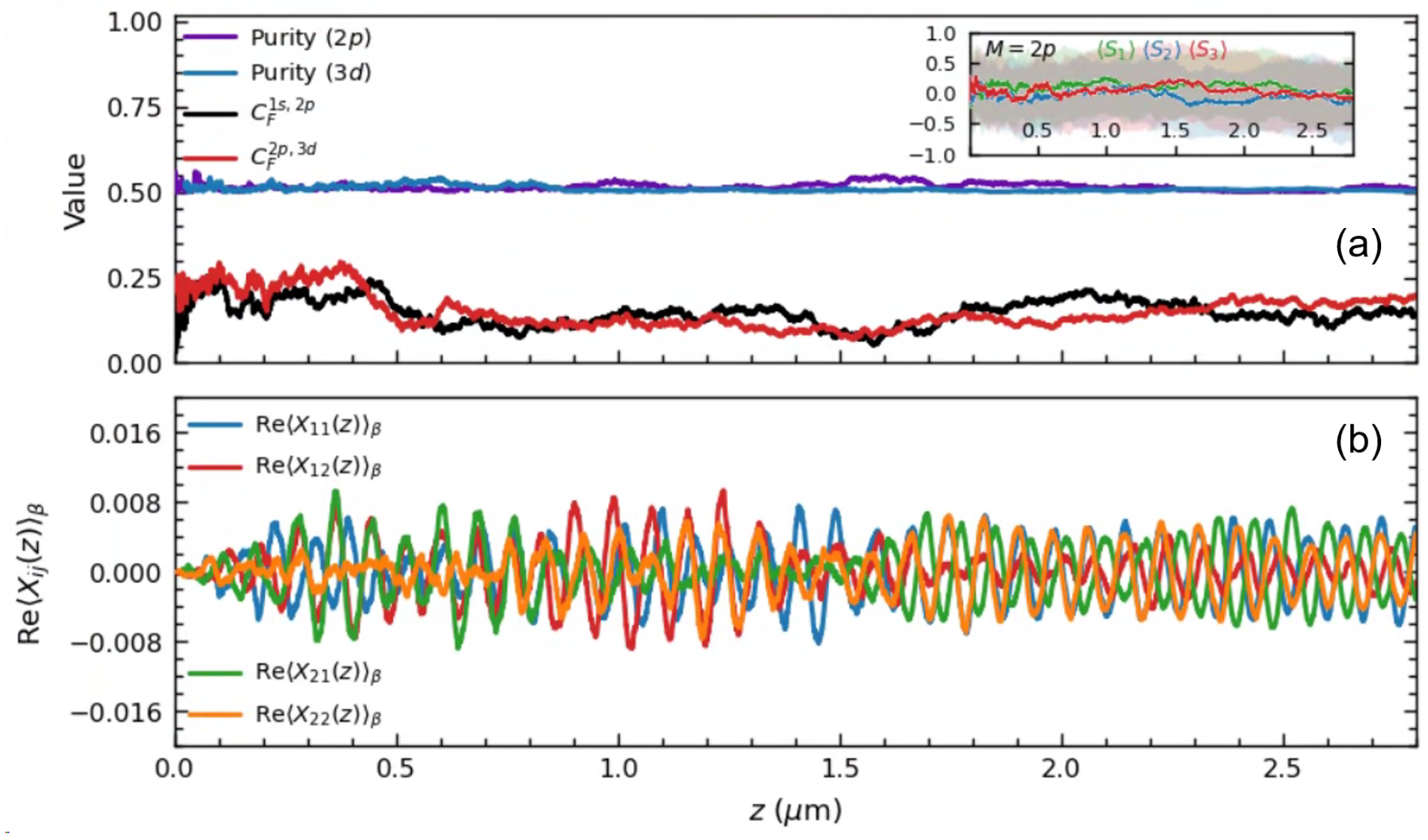

Fig. 4. (a) Intra-manifold purities $\gamma(z)$ for the degenerate $2p$ and $3d$ manifolds, together with Frobenius-normalized cross-coherences $C_F^{A,B}(z)$ for the manifold pairs $(1s, 2p)$ and $(2p, 3d)$, under $1s$ excitation. Inset shows configuration-averaged Bloch components $\langle S_i(z)\rangle_\beta$ within the $2p$ manifold for the same $1s$-initialized ensemble. (b) Population-unnormalized ensemble-averaged cross-block elements Re $\langle X_{ij}(z)\rangle_\beta$ for the $2p$–$3d$ cross block, where $i$ and $j$ index the chosen bases in the $2p$ and $3d$ manifolds, respectively.

## 4. Discussion

The comparison in Fig. 1 should be viewed primarily as a validation check. Consistent with established dynamical-diffraction and channeling-radiation treatments, the mean-field Debye–Waller potential recovers the thermally shifted transverse eigenenergies. The important distinction is that this static mean-field propagation does not generate configuration-dependent momentum transfer. The autocorrelation decay and bound-state population redistribution therefore require the frozen-phonon ensemble, where each displaced lattice realization produces a different transverse scattering operator.

The population-transfer results in Fig. 2 show that thermal diffuse scattering is strongly state selective. The preferential feeding among the $1s \leftrightarrow 2p$ and $2p \leftrightarrow 3d$ channels is governed by matrix elements of the thermal perturbation to the continuum string potential, $\delta V_\beta(r_\perp)$.

The intra-manifold results show that population transfer and coherence retention are distinct. The decay of $\gamma(z)$ toward 1/2 in the $2p$ manifold reflects stochastic mixing within the degenerate subspace, not merely population loss. Each frozen-phonon realization $\beta$ breaks the ideal axial symmetry of the string potential and defines a slightly different local eigenbasis inside the $m = \pm 1$ doublet. Since the thermal displacements are mean-zero and approximately isotropic, the ensemble average over these configuration-dependent mixing angles removes the macroscopic intra-manifold phase relation. The same mechanism explains why population transferred from $1s$ into the $2p$ and $3d$ manifolds is already close to maximally mixed within each degenerate subspace. The use of a common mean-field basis for all configurations allows for defining coherence in a fixed crystal-frame reference. A co-moving basis for each configuration would remove part of the realization-to-realization rotation that the ensemble average is intended to quantify.

The finite cross-manifold estimate in Fig. 4 does not establish that the 2p or 3d manifolds preserve a well-defined internal phase in a fixed basis. It is conceivable that amplitudes generated within a single realization may share common thermal fluctuations. The $2p$ and $3d$ components can therefore undergo partially correlated stochastic evolution. However, the small $C_F^{(2p,3d)}$ estimate is reported presently as an upper bound for a deterministic phase relation due to the possibility of a finite ensemble bias.

Coherence enters radiation observables through bilinear contractions of transition amplitudes $\mathcal{M}_{fm}$ with the electronic density matrix $\rho_{mn}$,

$$(26) \quad \frac{d^2N}{d\omega\, d\Omega} \propto \sum_{f,m,n} \rho_{mn}\,(z)\, \mathcal{M}_{fm}(\omega,\Omega)\, \mathcal{M}^*_{fn}(\omega,\Omega).$$

In this form, the loss of intra-manifold coherence in the $2p$ and $3d$ doublets directly suppresses interference terms that would otherwise encode azimuthal anisotropy and polarization structure in the corresponding channeling-radiation lines [28]. In the limit $\rho_{2p} \to I/2$, the radiation yield from the $2p$ manifold approaches the incoherent average over the degenerate subspace. A finite cross-manifold coherence is unlikely to be prominent in ordinary photon-yield spectra, where energy and angular averaging further suppress rapidly oscillating phase correlations. It may, however, become relevant in phase-sensitive settings, including externally driven or stimulated channeling-radiation schemes [15], where the relative phase between transverse amplitudes is part of the controlled dynamics.

## 5. Conclusions

Frozen-phonon multislice propagation was applied to MeV-range axial electron channeling, where the transverse bound-state spectrum is strongly quantized. The comparison with experimental channeling-radiation energies shows that the approach reproduces the relevant transverse transition energies, while also providing information that is not available from a static Debye–Waller potential.

By projecting the propagated wavefunction onto selected bound-state manifolds, the method gives direct access to ensemble-averaged manifold populations, reduced density matrices within degenerate subspaces, and cross-manifold coherence measures. For 16.9 MeV axial channeling in $\langle 100 \rangle$ diamond, thermal diffuse scattering was found to produce strongly state-selective population transfer. The dominant couplings occur between the $1s \leftrightarrow 2p$ and $2p \leftrightarrow 3d$ manifolds under the conditions studied. The coherence analysis shows that population transfer into a degenerate manifold does not imply preservation of a fixed internal phase relation. An initially pure $2p$ superposition evolves toward a nearly maximally mixed state due to configuration-dependent mixing within the degenerate subspace. Similarly, population transferred from an initial $1s$ excitation into the $2p$ and $3d$ manifolds is internally close to maximally mixed. The finite cross-manifold estimate provides an upper bound on persistent ensemble coherence.

These results show that frozen-phonon multislice propagation combined with bound-state projection provides a state-resolved description of thermal diffuse scattering in the strongly quantized channeling regime. The approach is particularly useful for studying thermal population dynamics, stochastic mixing of degenerate manifolds, and coherence-sensitive channeling-radiation observables. Because the wave-propagation framework is not restricted to ideal axial columns, it can be extended to more complex crystal geometries, including lattice defects, strained regions, and bent-crystal perturbations.

**Data Availability**

The data that support the findings of this article are not publicly available. The data are available from the author upon reasonable request.

**Declaration of competing interest**

The author declares that he has no known competing financial interests or personal relationships that could have appeared to influence the work reported in this paper.

# Supplemental Material

## Thermal Decoherence and Population Transfer of MeV Channeling Electrons in Diamond

Tadas Paulauskas

State Research Institute Center for Physical Sciences and Technology, Saulėtekio al. 3, Vilnius LT-10257, Lithuania

## Contents

### S1. Numerical implementation

The standard paraxial approximation for spinless high-energy channeling particles is used. Propagation optical $z$-axis plays the role of the evolution parameter for the transverse envelope $\psi(\boldsymbol{r}, z)$, which obeys Schrödinger-type equation

$$\text{(S1)} \quad \frac{i}{\sigma}\frac{\partial}{\partial z}\psi(\boldsymbol{r}, z) = \left[-\frac{\hbar^2}{2\gamma m_0}\nabla_{\boldsymbol{r}}^2 - V(\boldsymbol{r}, z)\right]\psi(\boldsymbol{r}, z).$$

Here $\gamma$ is Lorentz factor, $k_0$ is incident electron wavenumber, and the relativistic interaction constant used in the main text $\sigma = \frac{\gamma m_0}{k_0 \hbar^2}$ The operator in brackets defines the transverse Hamiltonian $H_{\perp}(z)$.

The high-energy electron wavefunction is propagated along the beam direction using a split-operator multislice scheme. A single propagation step of thickness $\Delta z$ is implemented as:

$$\text{(S2)} \quad \psi(x, y; z + \Delta z) = \mathcal{F}^{-1}\left[e^{i\kappa k_{\perp}^2 \Delta z}\, \mathcal{F}\left(e^{i\sigma V(x,y)\Delta z}\, \psi(x, y; z)\right)\right],$$

where $\mathcal{F}$ denotes the 2D transverse Fourier transform, $k_{\perp}^2 = k_x^2 + k_y^2$, $V(x, y)$ is the slice-projected electrostatic potential. The constant $\kappa$ equals $-1/(2k_0)$, the signed Fourier-space kinetic prefactor. The transverse frequency variables $(k_x, k_y)$ are the FFT angular spatial frequencies (units of $\text{Å}^{-1}$). To suppress FFT wrap-around artifacts, a smooth absorbing mask was applied near the outer $\sim 3\%$ of the supercell edges after each unit-cell propagation.

For ⟨100⟩ diamond, four equal-thickness slices per unit cell were used, while for ⟨110⟩ two slices per unit cell were employed, corresponding to step sizes $\Delta z \approx 0.89175$ Å and $1.26112$ Å, respectively. Besides the indicated propagator, the symmetric splitting:

(S3) $U_j \approx \exp[i\sigma V_j \Delta z/2] \exp[i\sigma K \Delta z] \exp[i\sigma V_j \Delta z/2]$,

was also tested and showed no observable differences in the numerical results reported in this work. Simulations used a $1024 \times 1024$ transverse grid on a $4 \times 4$ unit-cell supercell, with $\Delta x = \Delta y \approx 0.014$ Å. The Doyle–Turner parametrization of the atomic potential, $v_{\text{atom}}(r, B)$, was employed as commonly used in channeling-radiation simulations:

(S4) $v_{\text{atom}}(r, B) = \sum_{i=1}^{4} \frac{a_i}{b_i + B} \exp\left(\frac{-4\pi^2 r^2}{b_i + B}\right)$,

with an isotropic Debye–Waller (DW) factor $B = 8\pi^2 u_{2d}^2$. For frozen-phonon simulations, $u_{\text{2d}}$ is the primary input parameter controlling the thermal displacement amplitude. For each configuration $\beta$, the potential is constructed as $V^{(\beta)}(\boldsymbol{r}) = \sum_l v_{\text{atom}}\left(\boldsymbol{r} - \boldsymbol{R}_l^{(\beta)}\right)$.

Using Trotter propagation, the wavefunction autocorrelation $A(z)$ and its Fourier spectrum $S(k_z)$ are computed:

(S5) $A(z) = \langle \psi(\mathbf{r}, 0) | \psi(\mathbf{r}, z) \rangle$,

(S6) $S(k_z) = \int_0^Z A(z)\, e^{ik_z z}\, dz$.

Peaks in $|S(k_z)|$ occur at $k_{z,n} = \sigma \varepsilon_n$, enabling direct conversion between longitudinal spatial frequencies and transverse energies [12,13,24]. To extract an eigenmode associated with a chosen peak $k_{z,n}$, phase-compensated accumulation of the propagated wavefunction $\psi(\boldsymbol{r}, z_j)$ is performed over total $M$ slices to pick out the desired eigenmode,

(S7) $\phi_n(\boldsymbol{r}) \approx \sum_{j=1}^{M} \psi(\boldsymbol{r}, z_j)\, e^{ik_{z,n} z_j} \Delta z$.

For each slice, the projected potential was formed by summing atomic contributions at their projected transverse positions. Mean-field thermal smearing of the bulk lattice was incorporated through DW broadening of the atomic profiles. A total of $N_\beta = 30$ frozen-phonon configurations were used for the reported population and coherence metrics. For population-only metrics, $N_\beta \approx 15$ already produced near-converged results. Convergence tests for the coherence-related metrics are shown in the next section.

For spectral analysis, configuration-resolved correlation functions were sampled once per unit cell along $z$ and Fourier transformed with respect to the discrete sampling coordinate. Spectra were formed for each configuration and then averaged incoherently over the ensemble. All spectra shown correspond to the same $N_\beta$ used for the underlying frozen-phonon ensemble.

The simulation code was developed in Python and executed on a desktop workstation. Typical runs required several hours to complete $N_\beta = 30$ frozen-phonon propagations, including evaluation of cross-correlations among approximately 4–6 transverse reference modes at each recorded depth. To reduce computational cost while preserving the dominant axial perturbations relevant for bound-state channeling, both full frozen-phonon displacement of the simulated lattice and a "central-column frozen-phonon" approximation were tested. The two approaches gave very close agreement for the reported population-transfer and coherence metrics within the numerical convergence of the frozen-phonon ensemble. The reported calculations therefore use the central-column frozen-phonon implementation. In this mode, explicit frozen-phonon displacements are applied only to atoms belonging to the selected axial column, whereas the surrounding lattice is retained as a mean-field, thermally smeared Debye–Waller potential. This hybrid model targets the leading effect of thermal fluctuations on channeled bound-state dynamics while avoiding the cost of displacing and re-projecting all atoms in the supercell for every slice.

As the beam energy increases and/or for heavier atomic species, the most strongly bound transverse states become more localized, tightening requirements on transverse grid resolution and on the fidelity of the atomic potential model. In such regimes, finer transverse sampling and/or more accurate atomic potential parameterizations beyond Doyle–Turner may be required to reliably capture both bound-state structure and the impact of small thermal displacements in frozen-phonon modeling. It may also be necessary to reduce $\Delta z$ further (e.g., via atomic-potential subslicing or sub-integration) to obtain highly converged transverse energies and eigenmodes. Hybrid schemes (e.g., central-column frozen-phonon with a Debye–Waller background) can substantially reduce computational cost while preserving the primary axial perturbations relevant for bound-state channeling. The independent frozen-phonon calculations can also be parallelized across CPU/GPU resources.

## S2. Configuration-number convergence of coherence metrics

The convergence of the coherence metrics is tested with respect to the number of frozen-phonon configurations $N_\beta$ used for ensemble averaging of intra-manifold purity, coherence, and cross-coherence metrics. Up to $N_\beta = 30$ configurations were computed for $\langle 100 \rangle$ diamond at 16.9 MeV. In general, $N_\beta \approx 25–30$ is required to obtain well-converged coherence-related metrics. For plotting, a subset consisting of $N_\beta$ configurations is selected. This choice can lead to sizable variations when only $\sim$10–15 configurations are included.

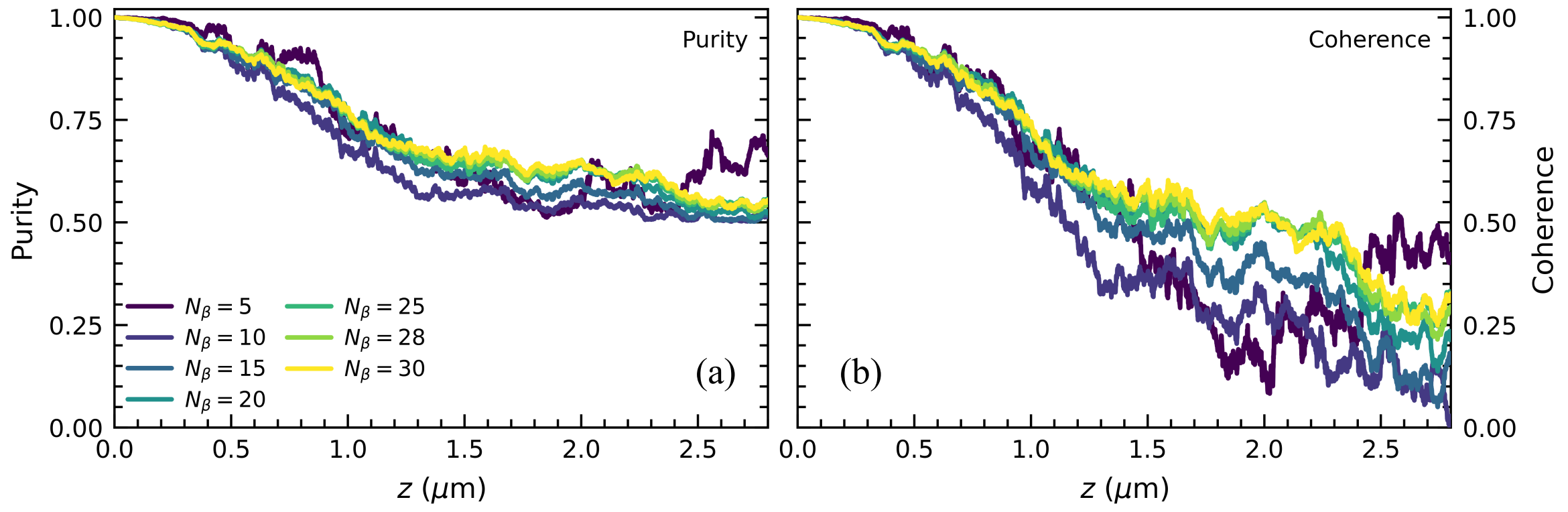

Fig. S1. Convergence test of (a) the purity and (b) the coherence of the degenerate $2p$ intra-manifold as a function of the number of frozen-phonon configurations $N_\beta$.

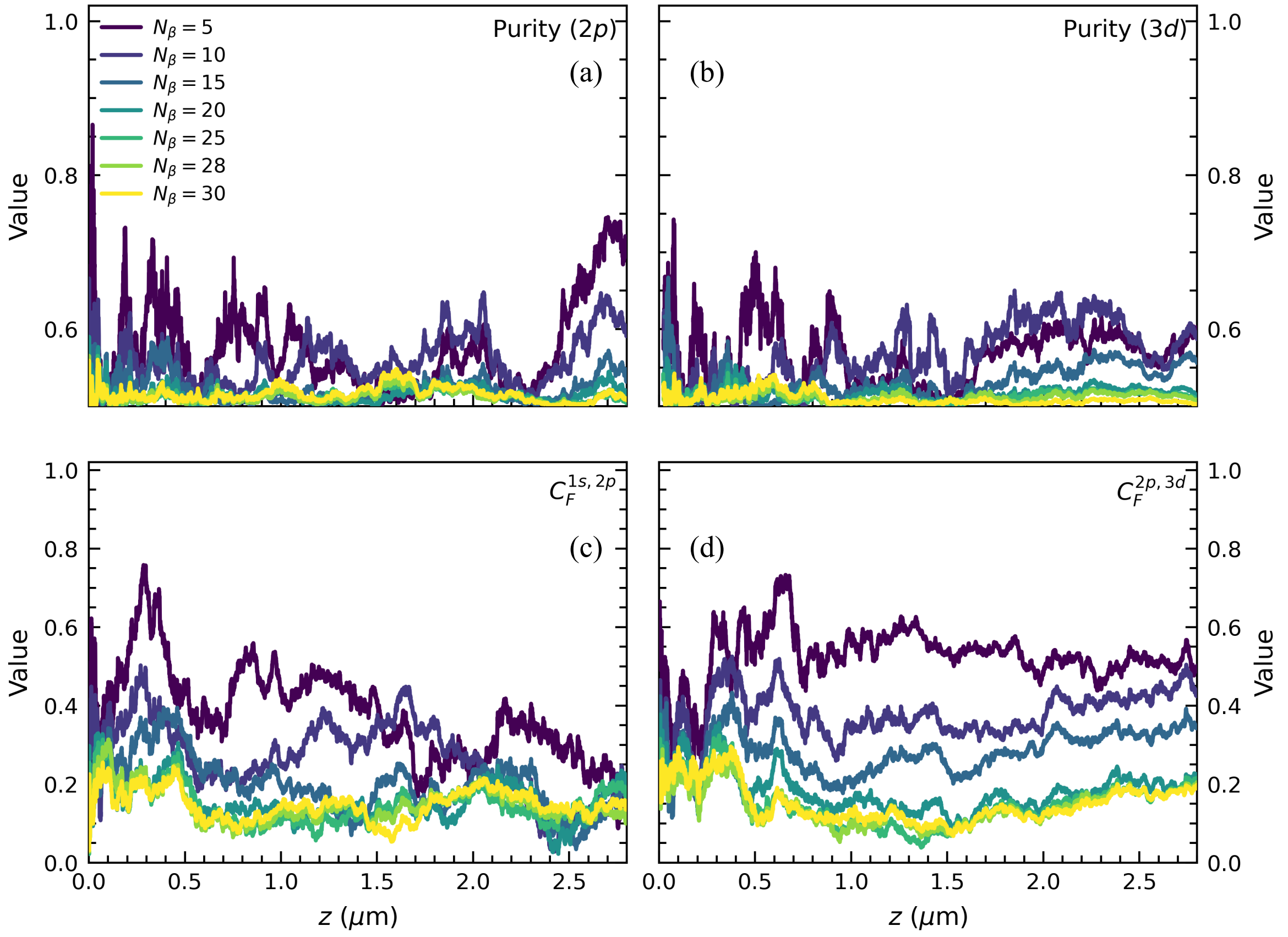

Fig. S2. Convergence test of (a,b) intra-manifold purity and (c,d) cross-coherence as a function of the number of frozen-phonon configurations $N_\beta$.

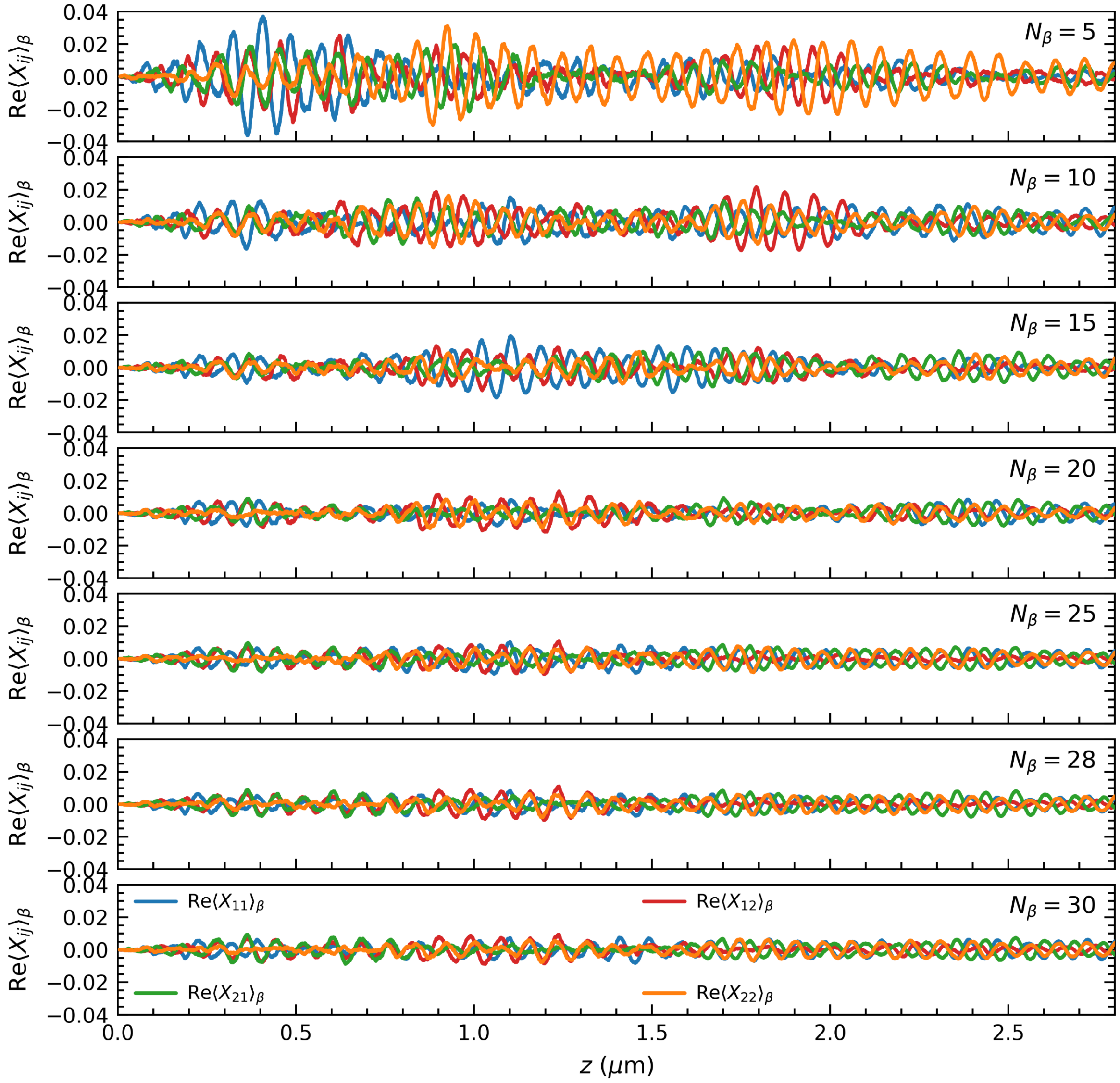

Fig. S3. Convergence test of the real part of the cross-block elements as a function of the number of frozen-phonon configurations $N_\beta$.

## S3. Slice-thickness convergence of 2*p* purity

The multislice discretization was tested with the same ten paired frozen-phonon configurations at 16.9 MeV and $u_{2d} = 0.055$ Å, using four and eight slices per unit cell. The long-depth mean purities are 0.5764 and 0.5758. The RMS and maximum absolute curve differences are $5.04\times10^{-4}$ and $1.34\times10^{-3}$, respectively. Thus, slice refinement has a negligible effect on the physical conclusion.

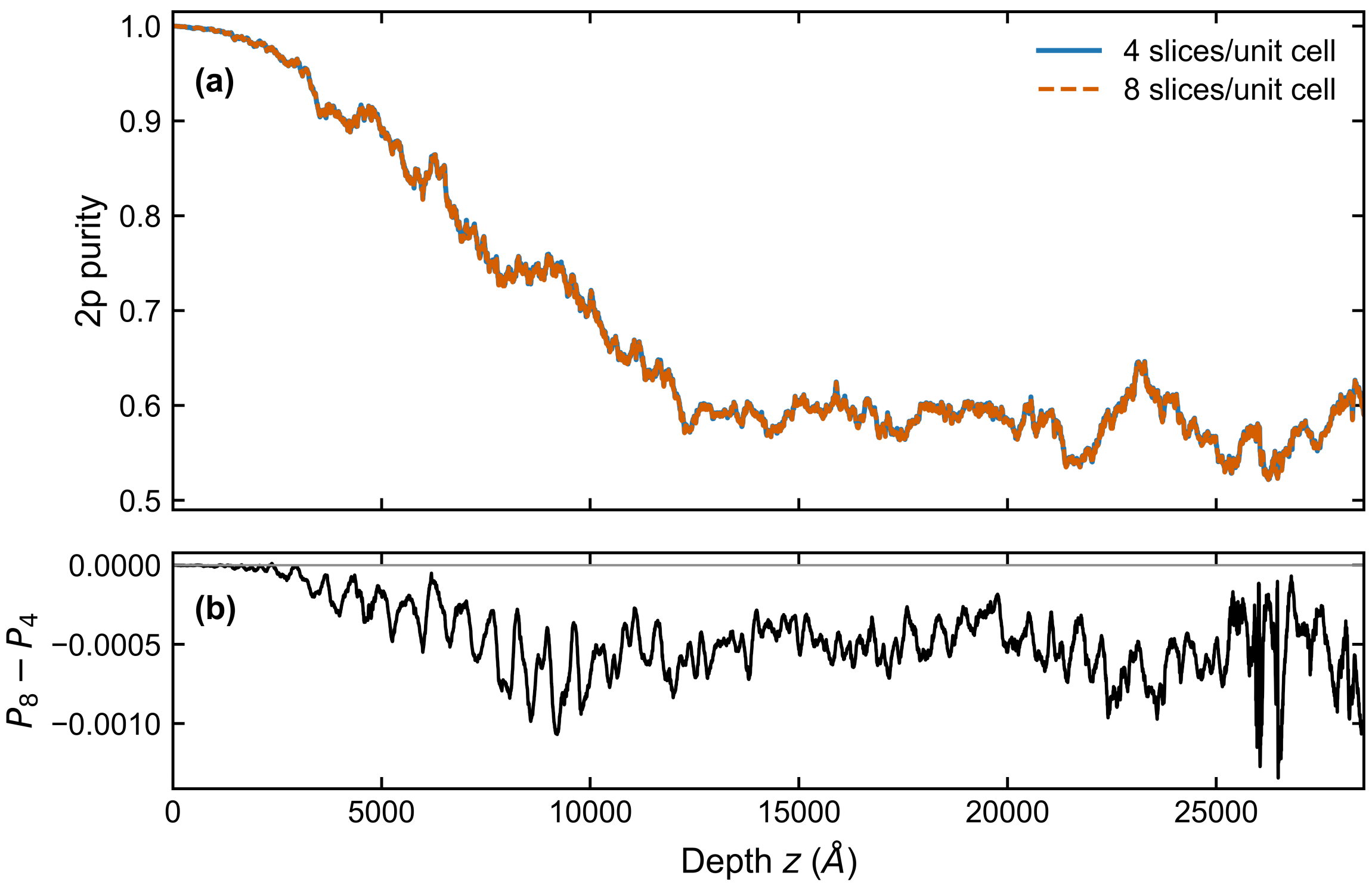

FIG. S4. Slice-thickness convergence of the 2p-subspace purity. (a) Purity for four and eight slices per unit cell using identical frozen-phonon configurations. (b) Pointwise difference $P_8 - P_4$. The overlapping curves and $10^{-3}$-scale difference show that the discretization is converged for this observable.

## S4. Thermal displacement amplitude sensitivity

Figure S5 shows mean-field 1*s*, 2*p*, 2*s*, and 3*d* transverse eigenvalues at 16.9 MeV extracted on the same discrete spectral grid for $u_{2d}$ = 0.040, 0.050, 0.055, and 0.070 Å. The energy-bin spacing is $\Delta E$ = 0.6947 eV. The 1*s* level moves across several bins and the 2*p* level across two bins over the full interval. The 2*s* level changes by one bin between 0.040 and 0.050 Å and is unresolved thereafter, whereas the 3*d* peak remains in the same assigned bin.

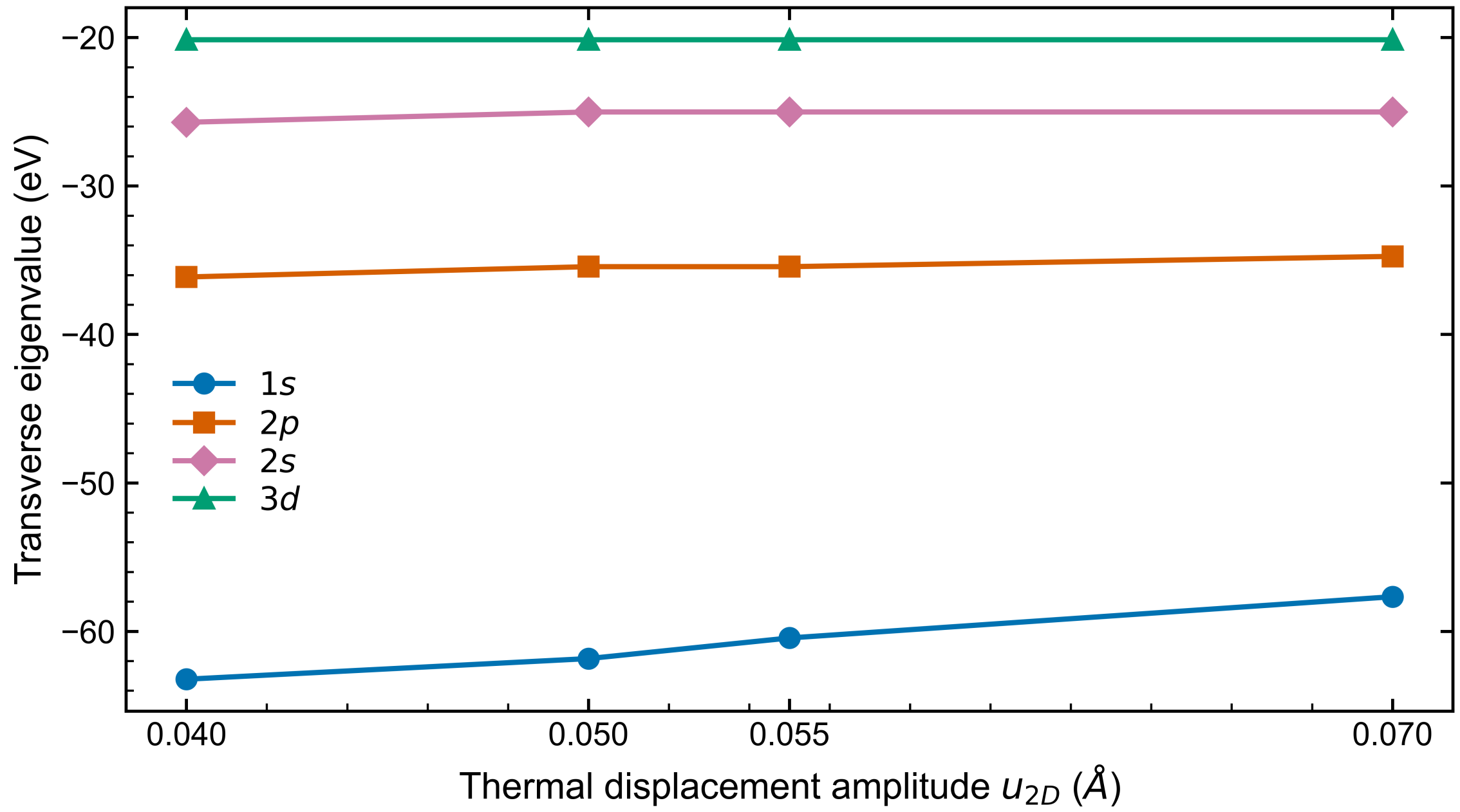

FIG. S5. Extracted 16.9 MeV transverse eigenvalues as a function of $u_{2d}$ for 0.040, 0.050, 0.055, and 0.070 Å

For the dynamical test, the same 30 frozen-phonon configurations were propagated at $u_{2d}$ = 0.040, 0.055, and 0.070 Å. The long-depth 2*p* purities are 0.682, 0.580, and 0.563, and the lower-amplitude case mixes more slowly. The quantitative rate is thermally sensitive, but substantial 2*p* mixing and the approach toward the maximally mixed limit persist across the tested range.

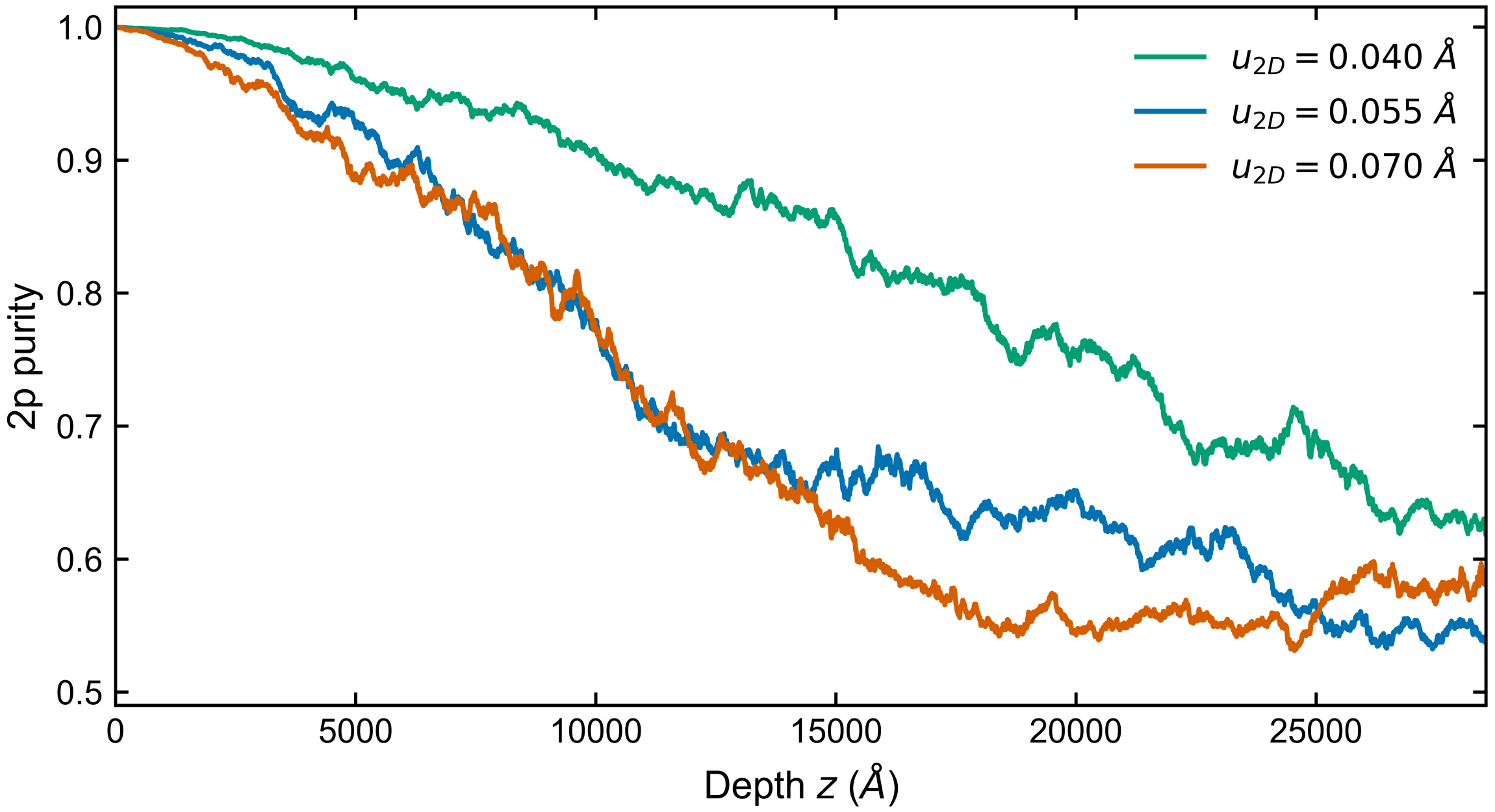

FIG. S6. 2*p*-subspace purity for $u_{2d}$ = 0.040, 0.055, and 0.070 Å at 16.9 MeV, calculated from the same thirty frozen-phonon configurations.

## S5. Statistical significance of cross-manifold coherence

The uncertainty of the ensemble cross-coherence was evaluated directly from the thirty configuration-resolved amplitudes. For the configuration bootstrap, thirty files were drawn repeatedly with replacement and the full ensemble cross-coherence was recalculated each time. The blue band in Fig. S7 therefore shows the sampling uncertainty expected for another ensemble of thirty comparable configurations. For the shuffled null, the $2p$ amplitudes from one configuration were randomly paired with the $3d$ amplitudes from another configuration. This deliberately breaks any genuine same-configuration link while preserving the individual amplitude distributions. The gray band is therefore the residual expected from finite sampling alone. Over 5000–25000 Å, the observed depth-averaged cross-coherence is 0.1765, with a configuration-bootstrap 95% interval of [0.1591, 0.3367]. The shuffled-null interval is [0.1377, 0.2301]. The maximum pointwise leave-one-out change is 0.0657, so no individual configuration dominates. Because the blue and gray ranges overlap, the finite estimate is not statistically distinguishable from the finite-ensemble background.

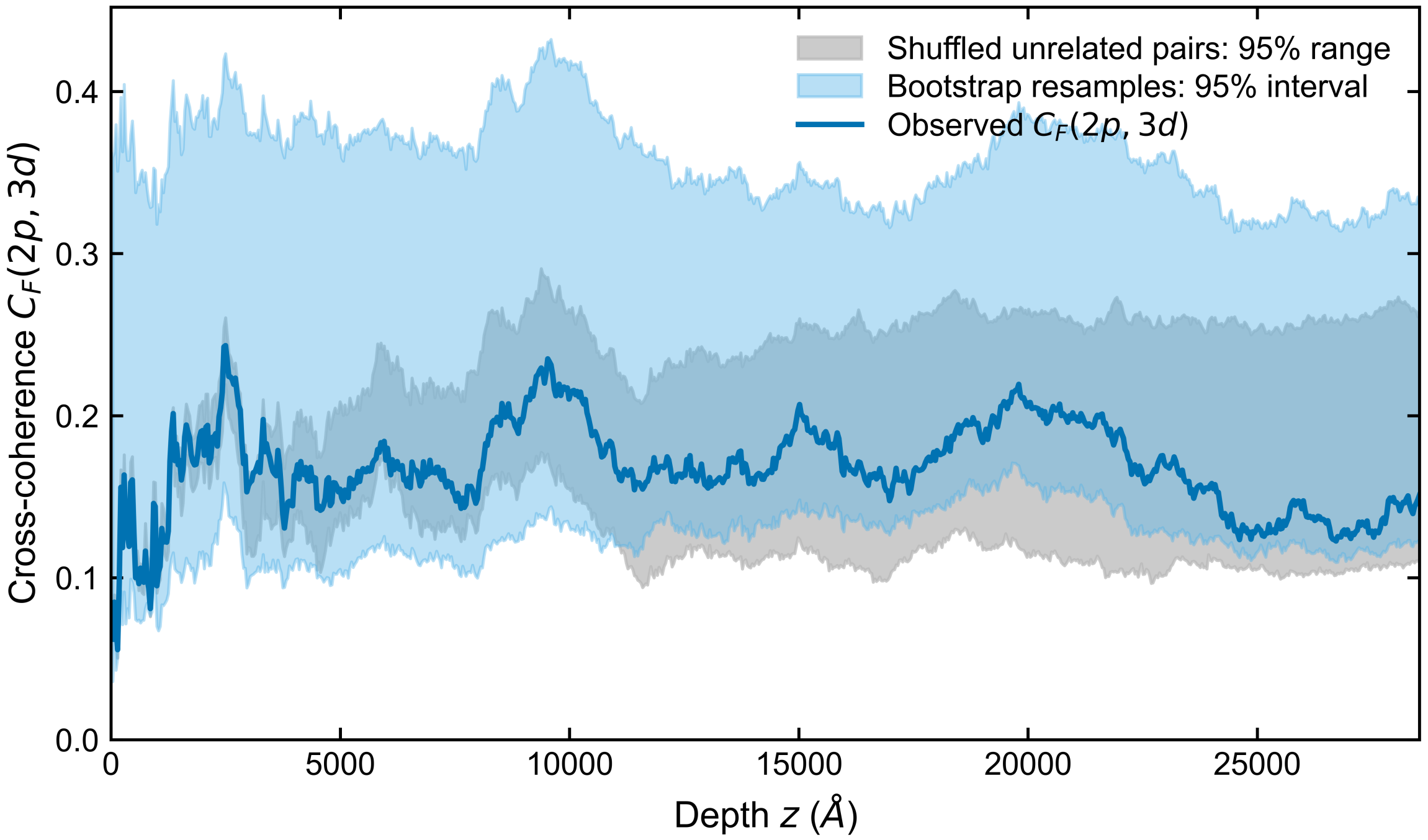

FIG. S7. Statistical assessment of the cross-manifold coherence. The line is the observed ensemble estimate. The blue band is the 95% interval obtained by repeatedly resampling the thirty configurations with replacement. The gray band is the 95% null range after pairing $2p$ and $3d$ amplitudes from unrelated configurations, thereby removing any same-configuration association. The trivial $z = 0$ normalization point is omitted.

## S6. Transverse bound-state eigenmodes

Figure S8 shows the transverse eigenmode probability densities $|\phi_n(\mathbf{r})|^2$ for the $1s, 2s, 2p,$ and $3d$ in $\langle 100\rangle$ diamond at 16.9 MeV. The states were extracted via spectral filtering in the full $4 \times 4$ unit-cell supercell using the mean-field (DW broadened) potential with $u_{2d} = 0.055$ Å. For the degenerate $2p$ manifold, the OAM basis states ($m = \pm 1$) have identical $|\phi(\mathbf{r})|^2$. For the $3d$ manifold, real-valued linear basis eigenmodes ($d_x$, $d_y$) are used and thus both are plotted. Images are shown on the native simulation grid (no interpolation). Pairwise orthogonality of the bound-states, computed via the inner products, gives residuals on the order of $10^{-6}$, which is adequate for the numerical accuracy required in the present channeling-dynamics calculations. The transverse energy of each extracted eigenmode $\phi_n$ was independently checked by evaluating the expectation value $\langle E_{\perp,n} \rangle$ within the mean-field supercell structure. It is the average over $L$-number of distinct slices (with transverse slice Hamiltonians $H_{\perp,g}$) in the repeating crystal unit cell along the optic axis:

$$\text{(S8)} \quad \langle E_{\perp,n} \rangle = \frac{1}{L} \sum_{g=1}^{L} \langle \phi_n | H_{\perp,g} | \phi_n \rangle.$$

Bound-state extraction is most reliable for strongly localized states. Multi-column hybridization in the periodic supercell can complicate identification. In full-crystal filtering, the selected column should be preferentially excited, otherwise states from multiple columns are extracted and may couple via scattering. An alternative is to extract states using an isolated-column potential. This, however, requires ensuring that extracted bound-states transverse energy in a supercell structure is not affected by the truncated structure, e.g., the lack of the mean inner-potential. Besides filtering, imaginary-z propagation (quantum-diffusion) can obtain the deepest bound states, with higher states generated by Gram–Schmidt orthogonalization.

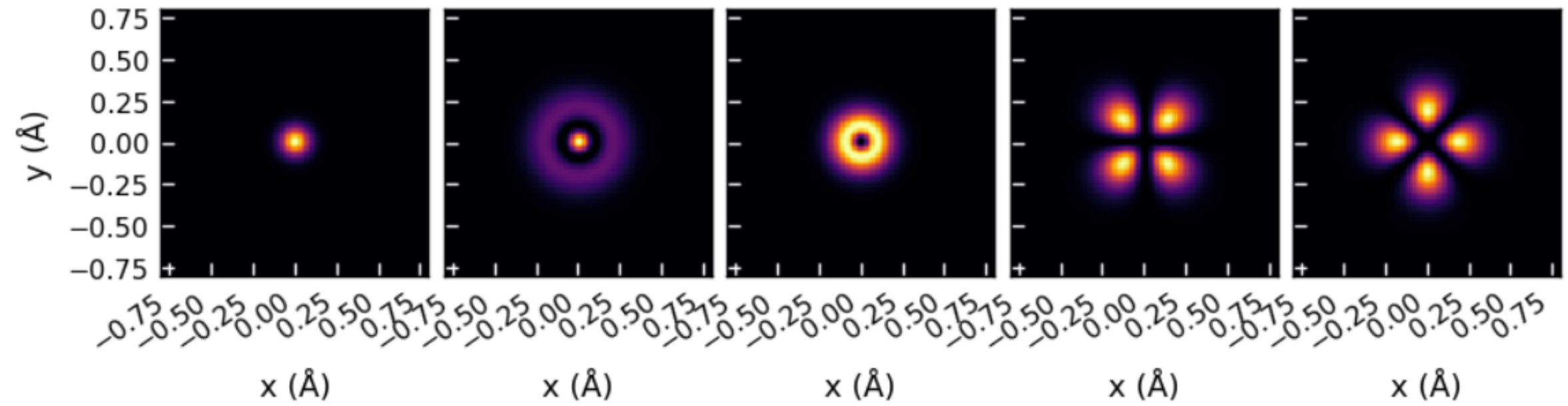

FIG. S8. Transverse bound-state probability densities $|\varphi_n(r)|^2$ for the $1s$, $2s$, $2p$, and $3d$ states, respectively. The OAM $2p$ states have identical intensity for m = ±1. For the $3d$ manifold, the two real-valued linear states $d_x$ and $d_y$ are shown.